\documentclass[titlepage,12pt]{article}
\usepackage{amssymb}
\usepackage{amsfonts}
\usepackage{amsmath}
\usepackage[dvips]{graphicx}
\usepackage[font=small,format=plain,labelfont=bf,up,textfont=it,up]{caption}

\begin{document}

\title{Transmission coefficient and two-fold degenerate discrete spectrum of
spin-1 bosons in a double-step potential\thanks {Preprint of an article submitted for consideration in International Journal
of Modern Physics E \copyright\ 2015, copyright World Scientific Publishing
Company.}}
\author{L. P. de Oliveira\thanks{%
luizp@if.usp.br} \\
%EndAName
Instituto de F\'{\i}sica,\\
Universidade de S\~{a}o Paulo (USP),\\
05508-900 S\~{a}o Paulo, SP, Brazil\\
and\\
A. S. de Castro\thanks{%
castro@pq.cnpq.br }\\
Departamento de F\'{\i}sica e Qu\'{\i}mica,\\
Universidade Estadual Paulista (UNESP),\\
12516-410 Guaratinguet\'{a}, SP, Brazil}
\date{}

\maketitle

\begin{abstract}
The scattering of spin-1 bosons in a nonminimal vector double-step potential
is described in terms of eigenstates of the helicity operator and it is
shown that the transmission coefficient is insensitive to the choice of the
polarization of the incident beam. Poles of the transmission amplitude
reveal the existence of a two-fold degenerate spectrum. The results are
interpreted in terms of solutions of two coupled effective Schr\"{o}dinger
equations for a finite square well with additional $\delta $-functions
situated at the borders. 

\bigskip


\bigskip


\bigskip

\noindent http://www.worldscientific.com/page/ijmpe
\end{abstract}

\section{Introduction}

The description of relativistic spin-1/2 and spin-0 particles embedded in a
double-step potential has been the subject of recent investigation and has
given rise to a striking result: bound-state solutions are obtained when the
potential is of sufficient intensity. As the potential is a double step, one
should not expect the existence of bound states, and it follows that such
bound states are consequence of the peculiar coupling. Spin-1/2 particles
were analyzed with a pseudoscalar coupling in the Dirac equation \cite{asc1}%
, whereas spinless particles were analyzed with a space component of a
nonminimal vector coupling \cite{asc2} in the context of the
Duffin-Kemmer-Petiau (DKP) formalism \cite{pet}-\cite{kem}.

The first-order DKP formalism describes spin-0 and spin-1 bosons with an
equation of motion which is of the same form for both kinds of bosons and
offers new horizons to introduce interactions in a straightforward way \cite%
{gue}-\cite{vij}. The nonminimal vector interaction is a kind of
charge-conjugate invariant coupling that behaves like a vector under a
Lorentz transformation. Because it does not couple to the charge of the
boson, it can not exhibit Klein's paradox \cite{ccc3}. Nonminimal vector
interactions, augmented by other kinds of Lorentz structures, have already
been used successfully in a phenomenological context to describe the
scattering of mesons by nuclei \cite{cla1}. Nonminimal vector couplings with
diverse functional forms for the potential functions have been explored in
the literature \cite{ccc3}, \cite{Ait}.

As found in Ref. \cite{asc2}, an interesting point regarding the scattering
of spin-0 bosons in a nonminimal vector double-step potential is that the
transmission coefficient never vanishes, regardless of the size of the
potential step. In sharp contrast with a nonrelativistic scheme, the
transmission amplitude exhibits complex poles corresponding to bound-state
solutions for a potential of sufficient intensity. Those results for
spinless bosons provide enough motivation for studying spin-1 bosons. In
this work we extend the analyses of Ref. \cite{asc2} in order to describe
the scattering of spin-1 bosons in terms of eigenstates of the helicity
operator. It is shown that the transmission coefficient, with the very same
form as that for spinless bosons, is insensitive to the choice of
polarization. The transmission amplitude exhibits poles corresponding to
bound-state solutions but, unlike spin-0 bosons, no zero-mode solution is
allowed. Furthermore, the discrete spectrum is two-fold degenerate with
respect to the helicity quantum numbers. The results are interpreted in
terms of solutions of two coupled effective Schr\"{o}dinger equations for a
finite square well with additional $\delta $-functions situated at the
borders.

The paper is organized as follows. In Sec. 2 we present the general features
of the DKP formalism with a nonminimal vector potential with emphasis on the
spin-1 sector of the theory. The spin-0 sector is drawn on succinctly in an
Appendix just for comparison with Ref. \cite{asc2}. In the subsequent
section we approach the scattering (Sec. 3.1) and bound-state solutions
(Sec. 3.2) in a double-step potential. In Sec. 4 we present our conclusions
and discuss how to use the helicity eigenvalue to get the solution for the
spin-0 bosons from the solution for spin-1 bosons.

\section{The space component of a nonminimal vector}

The DKP equation for a free boson is given by \cite{kem} (with units in
which $\hbar =c=1$)%
\begin{equation}
\left( i\beta ^{\mu }\partial _{\mu }-m\right) \psi =0  \label{dkp}
\end{equation}%
\noindent where the matrices $\beta ^{\mu }$ satisfy the algebra $\beta
^{\mu }\beta ^{\nu }\beta ^{\lambda }+\beta ^{\lambda }\beta ^{\nu }\beta
^{\mu }=g^{\mu \nu }\beta ^{\lambda }+g^{\lambda \nu }\beta ^{\mu }$
\noindent and the metric tensor is $g^{\mu \nu }=\,$diag$\,(1,-1,-1,-1)$.
That algebra generates a set of 126 independent matrices whose irreducible
representations are a trivial representation, a five-dimensional
representation describing the spin-0 particles and a ten-dimensional
representation associated with spin-1 particles. The second-order
Klein-Gordon and Proca equations are obtained when one selects the spin-0
and spin-1 sectors of the DKP theory. A conserved four-current is given by $%
J^{\mu }=\bar{\psi}\beta ^{\mu }\psi $, where the adjoint spinor $\bar{\psi}$
is given by $\bar{\psi}=\psi ^{\dagger }\eta ^{0}$ with $\eta ^{0}=2\left(
\beta ^{0}\right) ^{2}-1$ in such a way that $\left( \eta ^{0}\beta ^{\mu
}\right) ^{\dagger }=\eta ^{0}\beta ^{\mu }$ (the matrices $\beta ^{\mu }$
are Hermitian with respect to $\eta ^{0}$). The time component of this
current is not positive definite but it may be interpreted as a charge
density. Then the normalization condition $\int d\tau \,J^{0}=\pm 1$ where
the plus (minus) sign must be used for a positive (negative) charge, and the
expectation value of any observable $\mathcal{O}$ may be given by
\begin{equation}
\left\langle \mathcal{O}\right\rangle =\frac{\int d\tau \,\bar{\psi}\beta
^{0}\mathcal{O}\psi }{\int d\tau \,\bar{\psi}\beta ^{0}\psi }  \label{exp}
\end{equation}%
where $\mathcal{O}$ must be Hermitian with respect to $\beta ^{0}$, namely $%
\left( \beta ^{0}\mathcal{O}\right) ^{\dagger }=\beta ^{0}\mathcal{O}$, for
insuring that $\left\langle \mathcal{O}\right\rangle $ is a real quantity.
As a consequence of covariance under rotation, the matrix defined as
\begin{equation}
S_{i}=i\varepsilon _{ijk}\beta ^{j}\beta ^{k}
\end{equation}%
satisfies the angular momentum algebra. The spin is automatically
incorporated into the DKP theory as a necessary ingredient for describing a
mixing of the various components of the spinor. Notice that $\left( \beta
^{0}S_{i}\right) ^{\dagger }=\left( \beta ^{0}S_{i}\right) $ so that $S_{i%
\text{ }}$has real expectation values and because $S_{i\text{ }}^{3}=S_{i%
\text{ }}$their eigenvalues are $0$ and $\pm 1$. By considering only the
space component of a nonminimal vector term, the DKP equation can be written
as%
\begin{equation}
\left( i\beta ^{\mu }\partial _{\mu }-m-i[P,\beta ^{i}]A_{i}\right) \psi =0
\label{dkp2}
\end{equation}%
where $P$ is a matrix that makes $\bar{\psi}[P,\beta ^{i}]\psi $ to behave
like a space component of a vector under a Lorentz transformation. Once
again $\partial _{\mu }J^{\mu }=0$ \cite{ccc3}. If the potential is
time-independent one can write $\psi (\vec{r},t)=\phi (\vec{r})\exp (-iEt)$,
where $E\in
%TCIMACRO{\U{211d} }%
%BeginExpansion
\mathbb{R}
%EndExpansion
$, in such a way that the time-independent DKP equation becomes%
\begin{equation}
\left[ \beta ^{0}E+i\beta ^{i}\partial _{i}-\left( m+i[P,\beta
^{i}]A_{i}\right) \right] \phi =0  \label{DKP10}
\end{equation}%
In this case $J^{\mu }=\bar{\phi}\beta ^{\mu }\phi $ and so the spinor $\phi
$ describes a stationary state.

For the case of spin 1, we choose a representation for the $\beta ^{\mu }$
matrices given by \cite{ned2}%
\begin{equation}
\beta ^{0}=%
\begin{pmatrix}
0 & \overline{0} & \overline{0} & \overline{0} \\
\overline{0}^{T} & \mathbf{0} & \mathbf{I} & \mathbf{0} \\
\overline{0}^{T} & \mathbf{I} & \mathbf{0} & \mathbf{0} \\
\overline{0}^{T} & \mathbf{0} & \mathbf{0} & \mathbf{0}%
\end{pmatrix}%
,\quad \beta ^{i}=%
\begin{pmatrix}
0 & \overline{0} & e_{i} & \overline{0} \\
\overline{0}^{T} & \mathbf{0} & \mathbf{0} & -is_{i} \\
-e_{i}^{T} & \mathbf{0} & \mathbf{0} & \mathbf{0} \\
\overline{0}^{T} & -is_{i} & \mathbf{0} & \mathbf{0}%
\end{pmatrix}
\label{betaspin1}
\end{equation}%
\noindent where $s_{i}$ are the 3$\times $3 spin-1 matrices, $e_{i}$ are the
1$\times $3 matrices $\left( e_{i}\right) _{1j}=\delta _{ij}$ and $\overline{%
0}=%
\begin{pmatrix}
0 & 0 & 0%
\end{pmatrix}%
$, while $\mathbf{I}$ and $\mathbf{0}$ designate the 3$\times $3 unit and
zero matrices, respectively. The spin matrix is given by%
\begin{equation}
S_{i}=%
\begin{pmatrix}
0 & \overline{0} & \overline{0} & \overline{0} \\
\overline{0}^{T} & s_{i} & \mathbf{0} & \mathbf{0} \\
\overline{0}^{T} & \mathbf{0} & s_{i} & \mathbf{0} \\
\overline{0}^{T} & \mathbf{0} & \mathbf{0} & s_{i}%
\end{pmatrix}%
,\quad i=1,2,3  \label{s1}
\end{equation}%
That $\vec{S}^{\,2}$ is a conserved quantity with $\left\langle \vec{S}%
^{\,2}\right\rangle =2$, as expected for spin-1 bosons, follows from the
fact that $\beta ^{0}\vec{S}^{\,2}=2\beta ^{0}$. In this representation $%
P=\,\beta ^{\mu }\beta _{\mu }-2=\mathrm{diag}\,(1,1,1,1,0,0,0,0,0,0)$ \cite%
{vij}. \noindent With the spinor written as $\psi ^{T}=\left( \psi
_{1},...,\psi _{10}\right) $, and partitioned as%
\begin{equation*}
\psi _{I}^{\left( +\right) }=\left(
\begin{array}{c}
\psi _{3} \\
\psi _{4}%
\end{array}%
\right) ,\quad \psi _{I}^{\left( -\right) }=\psi _{5}
\end{equation*}%
\begin{equation}
\psi _{II}^{\left( +\right) }=\left(
\begin{array}{c}
\psi _{6} \\
\psi _{7}%
\end{array}%
\right) ,\quad \psi _{II}^{\left( -\right) }=\psi _{2}  \label{part}
\end{equation}%
\begin{equation*}
\psi _{III}^{\left( +\right) }=\left(
\begin{array}{c}
\psi _{10} \\
-\psi _{9}%
\end{array}%
\right) ,\quad \psi _{III}^{\left( -\right) }=\psi _{1}
\end{equation*}%
the DKP equation for a boson constrained to move along the $X$-axis
decomposes into
\begin{equation*}
\left( \partial _{0}^{2}-D_{1}^{\left( -\sigma \right) }D_{1}^{\left( \sigma
\right) }+m^{2}\right) \psi _{I}^{\left( \sigma \right) }=0
\end{equation*}%
\begin{equation}
\psi _{II}^{\left( \sigma \right) }=\frac{i}{m}\,\partial _{0}\psi
_{I}^{\left( \sigma \right) },\quad \psi _{III}^{\left( \sigma \right) }=%
\frac{i}{m}\,D_{1}^{\left( \sigma \right) }\psi _{I}^{\left( \sigma \right) }
\label{DKp3}
\end{equation}%
\begin{equation*}
\psi _{8}=0
\end{equation*}%
where $\sigma $ is equal to $+$ or $-$, and%
\begin{equation}
D_{1}^{\left( \sigma \right) }=\partial _{1}+\sigma A_{1}
\end{equation}%
Before continuing, note that the third line plus the second equation in the
middle line of (\ref{DKp3}) are the constraint equations which allow one to
eliminate the superfluous components ($\psi _{1}$, $\psi _{8}$, $\psi _{9}$
and $\psi _{10}$) of the DKP spinor. The component $\psi _{8}=0$ because the
movement is restrict to the $X$-axis. Because $S_{1}$ commutes with $\beta
^{0},\beta ^{1}$ and $[P,\beta ^{1}]$, one can envisage another conserved
quantity beyond $\vec{S}^{\,2}$: the projection of the spin onto the
direction of the motion (helicity operator). Indeed,
\begin{equation}
\left\langle S_{1}\right\rangle =\frac{\int_{-\infty }^{+\infty }dx\,\text{Re%
}\left( \psi _{4}^{\ast }\,\partial _{0}\psi _{3}-\psi _{3}^{\ast
}\,\partial _{0}\psi _{4}\right) }{\int_{-\infty }^{+\infty }dx\,\text{Im}%
\left( \sum\limits_{\sigma }\psi _{I}^{\left( \sigma \right) \dagger
}\partial _{0}\psi _{I}^{\left( \sigma \right) }\right) }
\end{equation}%
measures the degree and direction of the polarization in such a way that $%
\psi _{I}^{\left( -\right) }$ is closely related to an eigenstate of the
helicity operator with expectation value $0$ ($\psi _{3}=\psi _{4}=0$), and $%
\psi _{I}^{\left( +\right) }$ with eigenvalues $\pm 1$ ($\psi _{4}=\pm i\psi
_{3}$ and $\psi _{5}=0$). Therefore, $\sigma $ plays an important role to
specify the states of polarization: $\sigma =-$ for transverse and $\sigma
=+ $ for longitudinal. The constancy of $\left\langle S_{1}\right\rangle $
implies that the degree of polarization of a beam of bosons entering a
region with a potential is not affected, no matter how complicated the space
component of the nonminimal vector potential may be.

Meanwhile the time-independent DKP equation can be written in the simpler
form
\begin{equation*}
\left( \frac{d^{2}}{dx^{2}}+k_{\sigma }^{2}\right) \phi _{I}^{\left( \sigma
\right) }=0
\end{equation*}%
\begin{equation}
\phi _{II}^{\left( \sigma \right) }=\frac{E}{m}\,\phi _{I}^{\left( \sigma
\right) }  \label{spin1-ti}
\end{equation}%
\begin{equation*}
\phi _{III}^{\left( \sigma \right) }=\frac{i}{m}\left( \frac{d}{dx}+\sigma
A_{1}\right) \phi _{I}^{\left( \sigma \right) },\quad \phi _{8}=0
\end{equation*}%
where%
\begin{equation}
k_{\sigma }^{2}=E^{2}-m^{2}-A_{1}^{2}+\sigma \frac{dA_{1}}{dx}
\end{equation}%
Now the components of the four-current are%
\begin{equation}
J^{0}=\frac{E}{m}\sum\limits_{\sigma }|\phi _{I}^{\left( \sigma \right)
}|^{2},\quad J^{1}=\frac{1}{m}\,\text{Im}\sum\limits_{\sigma }\phi
_{I}^{\left( \sigma \right) \dagger }\,\frac{d\phi _{I}^{\left( \sigma
\right) }}{dx},\quad J^{2}=J^{3}=0  \label{CUR2}
\end{equation}

Given that the interaction potential satisfies certain conditions, we have a
well-defined Sturm-Liouville problem and hence a natural and definite method
to investigate stationary states with continuous or discrete eigenvalues.

\section{The double-step potential}

\noindent Let us now consider the double-step potential
\begin{equation}
A_{1}\left( x\right) =V_{0}\left[ \theta \left( x-a\right) -\theta \left(
-x-a\right) \right]  \label{sc1a}
\end{equation}%
with $V_{0}$ and $a$ defined to be real numbers ($a>0$) and $\theta \left(
x\right) $ is the Heaviside step function. The double-step function reduces
to the sign function when $a=0$. With (\ref{sc1a}) the first line of (\ref%
{spin1-ti}) can be written as%
\begin{equation*}
\frac{d^{2}\phi _{I}^{\left( \sigma \right) }}{dx^{2}}+\left\{
E^{2}-m^{2}+\sigma V_{0}\left[ \delta \left( x-a\right) +\delta \left(
x+a\right) \right] \right.
\end{equation*}%
\begin{equation}
\qquad \qquad \qquad \qquad \qquad \qquad \left. -V_{0}^{2}\left[ \theta
\left( x-a\right) +\theta \left( -x-a\right) \right] \right\} \phi
_{I}^{\left( \sigma \right) }=0  \label{eq1}
\end{equation}%
where $\delta \left( x\right) =d\theta \left( x\right) /dx$ is the Dirac
delta function.

\subsection{Scattering}

We turn our attention to states belonging to the continuous spectrum ($%
E^{2}>m^{2}+V_{0}^{2}$) corresponding to a physical situation in which
bosons with a given energy and polarization impinge on the potential. Our
problem is to describe, after interacting, the way the bosons move off. The
solutions describing bosons coming from the left can be written as
\begin{equation}
\phi _{I}^{\left( \sigma \right) }(x)=\left\{
\begin{array}{cc}
A^{\left( \sigma \right) }e^{+i\eta \frac{x}{a}}+B^{\left( \sigma \right)
}e^{-i\eta \frac{x}{a}} & \text{\textrm{for }}x<-a \\
&  \\
C^{\left( \sigma \right) }e^{+i\xi \frac{x}{a}}+D^{\left( \sigma \right)
}e^{-i\xi \frac{x}{a}} & \text{\textrm{for }}|x|<a \\
&  \\
F^{\left( \sigma \right) }e^{+i\eta \frac{x}{a}} & \text{\textrm{for }}x>a%
\end{array}%
\right.  \label{eq50}
\end{equation}%
\noindent where%
\begin{equation}
\xi =a\sqrt{E^{2}-m^{2}},\quad \eta =\sqrt{\xi ^{2}-\mathcal{\upsilon }^{2}}%
,\quad \mathcal{\upsilon }=aV_{0}
\end{equation}%
\noindent Recall that $\sigma $ is related to either a longitudinally
polarized state ($\sigma =$ $+$) or a transversely polarized state ($\sigma
= $ $-$). $\phi _{I}^{\left( \sigma \right) }$ describes an incident wave
moving to the right ($\eta $ is a real number) and a reflected wave moving
to the left with%
\begin{equation}
J^{1}\left( x<-a\right) =\frac{\eta }{am}\sum\limits_{\sigma }\left(
|A^{\left( \sigma \right) }|^{2}-|B^{(\sigma )}|^{2}\right)
\end{equation}%
and a transmitted wave moving to the right with%
\begin{equation}
J^{1}\left( x>a\right) =\frac{\eta }{am}\sum\limits_{\sigma }|F^{\left(
\sigma \right) }|^{2}
\end{equation}%
We demand that $\phi _{I}^{\left( \sigma \right) }$ be continuous at $x=\pm
a $. Effects due to the potential on $d\phi _{I}^{\left( \sigma \right) }/dx$
in the neighbourhood of $x=\pm a$ can be evaluated by integrating (\ref{eq1}%
) from $\pm a-\varepsilon $ to $\pm a+\varepsilon $ and taking the limit $%
\varepsilon \rightarrow 0$. The connection formula between $d\phi
_{I}^{\left( \sigma \right) }/dx$ at the right and $d\phi _{I}^{\left(
\sigma \right) }/dx$ at the left can be summarized as%
\begin{equation}
\lim_{\varepsilon \rightarrow 0}\left. \frac{d\phi _{I}^{\left( \sigma
\right) }}{dx}\right\vert _{x=\pm a-\varepsilon }^{x=\pm a+\varepsilon }=-%
\frac{\sigma \mathcal{\upsilon }}{a}\,\phi _{I}^{\left( \sigma \right) }(\pm
a)  \label{discont}
\end{equation}%
Thus, one gets the relative amplitudes%
\begin{eqnarray}
\frac{B^{\left( \sigma \right) }}{A^{\left( \sigma \right) }} &=&\frac{%
2ie^{-2i\eta }\,\xi \sigma \mathcal{\upsilon }\cos 2\xi }{d^{(\sigma )}}
\notag \\
&&  \notag \\
\frac{C^{\left( \sigma \right) }}{A^{\left( \sigma \right) }} &=&\frac{%
e^{-i\xi }e^{-i\eta }\,\eta \left( \xi +\eta -i\sigma \mathcal{\upsilon }%
\right) }{d^{(\sigma )}}  \notag \\
&&  \label{amp} \\
\frac{D^{\left( \sigma \right) }}{A^{\left( \sigma \right) }} &=&\frac{%
e^{i\xi }e^{-i\eta }\,\eta \left( \xi -\eta +i\sigma \mathcal{\upsilon }%
\right) }{d^{(\sigma )}}  \notag \\
&&  \notag \\
\frac{F^{(\sigma )}}{A^{\left( \sigma \right) }} &=&\frac{2e^{-2i\eta }\,\xi
\eta }{d^{(\sigma )}}  \notag
\end{eqnarray}%
where%
\begin{equation}
d^{(\sigma )}\equiv 2\left( \eta -i\sigma \mathcal{\upsilon }\right) \left(
\xi \cos 2\xi -i\eta \sin 2\xi \right)
\end{equation}

In order to determinate the reflection and transmission coefficients we use
the charge current densities $J^{1}\left( x<-a\right) $ and $J^{1}\left(
x>a\right) $. The $x$-independent current density allow us to define the
transmission coefficients as%
\begin{equation}
T=\frac{\sum\limits_{\sigma }|F^{\left( \sigma \right) }|^{2}}{%
\sum\limits_{\sigma }|A^{\left( \sigma \right) }|^{2}}
\end{equation}%
The last relative amplitude in (\ref{amp}) allow us to write%
\begin{equation}
T=\left[ 1+\left( \frac{\mathcal{\upsilon }}{\eta }\cos 2\xi \right) ^{2}%
\right] ^{-1}
\end{equation}%
regardless of the sign of $\sigma \mathcal{\upsilon }$. Notice that $%
T\rightarrow 1$ as $\eta \rightarrow \infty $ and that there is a resonance
transmission ($T=1$) whenever $\xi =\left( 2n+1\right) \pi /4$ with $%
n=0,1,2,\ldots $

Note that it is possible to construct initial states representing bosons
polarized in any direction before the scattering ($A^{\left( +\right) }=0$
or $A^{\left( -\right) }=0$) and that the transmission amplitude for $\sigma
$-polarized bosons ($F^{(\sigma )}/A^{\left( \sigma \right) }$) is $\sigma $%
-dependent. Nevertheless, the transmission coefficient is not sensitive to
the exact choice of polarization. Comparison with the spin-0 case \cite{asc2}
shows that spin-1 and spin-0 particles have the same transmission
coefficient. Because of the $\sigma $-dependence of the transmission
amplitude, though, the poles of the transmission amplitude might reveal
differences in the bound-state spectra of spin-1 and spin-0 particles.

\subsection{Bound states}

As for bound states ($E^{2}<m^{2}+V_{0}^{2}$), our problem is to solve (\ref%
{eq1}) for $\phi ^{\left( \sigma \right) }$ and to determine the allowed
energies. The possibility of bound states requires a solution given by (\ref%
{eq50}) with $\eta =i|\eta |$ ($\xi <|\mathcal{\upsilon }|$) and $A^{\left(
\sigma \right) }=0$ in order to obtain a square-integrable spinor.
Therefore, if one considers the transmission amplitude as a function of the
complex variable $\eta $ one sees that for $\eta $ real and positive one
obtains the scattering states whereas the bound states would be obtained by
the poles lying along the positive imaginary axis of the complex $\eta $%
-plane. The equations for $\phi _{I}^{\left( \sigma \right) }$ and $\phi
_{I}^{\left( -\sigma \right) }$ are not independent because $E$ appears in
both equations and so one has to search for bound-state solutions for $\phi
_{I}^{\left( \sigma \right) }$ and $\phi _{I}^{\left( -\sigma \right) }$
with a common $E$. This means that all the spectrum of spin-1 bosons is
two-fold degenerate with respect to $\sigma $.

Setting $|\eta |=\sqrt{\mathcal{\upsilon }^{2}-\xi ^{2}}$, one sees that $%
F^{(\sigma )}/A^{\left( \sigma \right) }$ has a pole at $\xi =0$ when $%
\sigma \mathcal{\upsilon }>0$, and $F^{(-\sigma )}/A^{\left( -\sigma \right)
}$ when $\sigma \mathcal{\upsilon }<0$. Poles independent of the sign of $%
\sigma $ are solutions of the transcendental equation%
\begin{equation}
\tan 2\xi =-\frac{\xi }{|\eta |}
\end{equation}%
With the amplitudes given by (\ref{amp}) one obtains the ratios
\begin{eqnarray}
\frac{C^{\left( \sigma \right) }}{D^{\left( \sigma \right) }} &=&e^{-i\left(
2\xi +\arctan \frac{\xi }{|\eta |}\right) }  \notag \\
&&  \label{Amp} \\
\frac{B^{\left( \sigma \right) }}{F^{(\sigma )}} &=&\frac{\sigma \mathcal{%
\upsilon }}{|\eta |}\cos 2\xi  \notag
\end{eqnarray}%
for $\eta =i|\eta |$. It is true that the first line of (\ref{Amp})
furnishes $|C^{\left( \sigma \right) }|=|D^{\left( \sigma \right) }|$. It
has to be so since the charge current density $J^{1}$ vanishes in the region
$|x|>a$ whereas in the region $|x|<a$ it takes the form%
\begin{equation}
J^{1}=\left\{
\begin{array}{cc}
\frac{\xi }{am}\sum\limits_{\sigma }\left( |C^{\left( \sigma \right)
}|^{2}-|D^{\left( \sigma \right) }|^{2}\right) & \text{\textrm{for }}\xi
\text{ \textrm{real}} \\
&  \\
\frac{2|\xi |}{am}\text{\thinspace }\sum\limits_{\sigma }\text{\textrm{Im}}%
\left( C^{\left( \sigma \right) \ast }D^{\left( \sigma \right) }\right) &
\text{\textrm{for }}\xi \text{ \textrm{imaginary}}%
\end{array}%
\right.
\end{equation}%
Hence, one concludes that bound states are only possible if $|C^{\left(
\sigma \right) }|=|D^{\left( \sigma \right) }|$. Since $A_{1}(x)$ is
antisymmetric with respect to $x$, it follows that $\phi _{I}^{\left( \sigma
\right) }$ can be either even or odd. Hence,
\begin{equation}
\frac{C^{\left( \sigma \right) }}{D^{\left( \sigma \right) }}=\frac{%
B^{\left( \sigma \right) }}{F^{(\sigma )}}=\pm 1  \label{ratio}
\end{equation}%
so that%
\begin{equation}
\phi _{I}^{\left( \sigma \right) }\left( x\right) =\left\{
\begin{array}{cc}
+F^{(\sigma )}e^{+|\eta |\frac{x}{a}} & \text{\textrm{for }}x<-a \\
&  \\
2C^{\left( \sigma \right) }\cos \left( \xi \frac{x}{a}\right) & \text{%
\textrm{for }}|x|<a \\
&  \\
+F^{(\sigma )}e^{-|\eta |\frac{x}{a}} & \text{\textrm{for }}x>+a%
\end{array}%
\right.
\end{equation}%
for $\phi _{I}^{\left( \sigma \right) }\left( -x\right) =+\phi _{I}^{\left(
\sigma \right) }\left( x\right) $, and%
\begin{equation}
\phi _{I}^{\left( \sigma \right) }\left( x\right) =\left\{
\begin{array}{cc}
-F^{(\sigma )}e^{+|\eta |\frac{x}{a}} & \text{\textrm{for }}x<-a \\
&  \\
2iC^{\left( \sigma \right) }\sin \left( \xi \frac{x}{a}\right) & \text{%
\textrm{for }}|x|<a \\
&  \\
+F^{(\sigma )}e^{-|\eta |\frac{x}{a}} & \text{\textrm{for }}x>+a%
\end{array}%
\right.
\end{equation}%
for $\phi _{I}^{\left( \sigma \right) }\left( -x\right) =-\phi _{I}^{\left(
\sigma \right) }\left( x\right) $. The condition $C^{\left( \sigma \right)
}/D^{\left( \sigma \right) }=\pm 1$ demands%
\begin{equation}
\tan 2\xi =-\frac{\xi }{|\eta |}
\end{equation}%
One can write this last relation as%
\begin{equation}
\xi \tan \xi =|\eta |-\lambda |\mathcal{\upsilon }|,\quad \lambda =\pm 1
\label{bs}
\end{equation}%
In addition, one may check that (\ref{bs}) implies into%
\begin{equation}
\frac{B^{\left( \sigma \right) }}{F^{(\sigma )}}=\lambda \,\mathrm{sgn}%
\left( \sigma \mathcal{\upsilon }\right)
\end{equation}%
Now we see that $\lambda \,$\textrm{sgn}$\left( \sigma \mathcal{\upsilon }%
\right) $ is the parity eigenvalue and that the equation for even-parity
solutions is mapped into that one for odd-parity solutions under the change
of $\sigma \mathcal{\upsilon }$ by $-\sigma \mathcal{\upsilon }$, and vice
versa. In addition, because $|\eta |=\sqrt{\mathcal{\upsilon }^{2}-\xi ^{2}}$%
, (\ref{bs}) can also be written as%
\begin{equation}
-\xi \cot \xi =|\eta |+\lambda |\mathcal{\upsilon }|,\quad \lambda =\pm 1
\end{equation}%
For $\xi =i|\xi |$ ($|\eta |=\sqrt{\mathcal{\upsilon }^{2}+|\xi |^{2}}$),
equation (\ref{bs}) is transformed into%
\begin{equation}
-|\xi |\tanh |\xi |=|\eta |-\lambda |\mathcal{\upsilon }|,\quad \lambda =\pm
1  \label{b2}
\end{equation}

For $\xi \in \,$$\mathbb{R}$ the solutions for bound states are given by the
intersection of the curve represented by $\xi \tan \xi $ with the curves
represented by $|\eta |-\lambda |\mathcal{\upsilon }|$. One can see that it
needs critical values, corresponding to $|\mathcal{\upsilon }_{\text{c}%
}|=-\arctan \lambda $, for obtaining bound states. If $a|V_{0}|$ is larger
than the critical values there will be a finite sequence of bound states
with alternating parities expressed as $\lambda \,$\textrm{sgn}$\left(
\sigma V_{0}\right) $. On the other hand, the left-hand side of (\ref{b2})
is always negative whereas its right-hand side is always positive and so (%
\ref{b2}) furnish no solutions.

Finally, in the limit $a\rightarrow 0$ ($\xi \rightarrow 0$) one has%
\begin{equation}
T\underset{a\rightarrow 0}{\longrightarrow }\left( 1+\frac{V_{0}^{2}}{%
E^{2}-m^{2}-V_{0}^{2}}\right) ^{-1}
\end{equation}%
and the transmission amplitude has no pole common to the cases $\sigma =+$
and $\sigma =-$. It is instructive to note that the quantization condition
for spin-1 particles is formally identical to that one for spin-0 particles
\cite{asc2} but the parity eigenvalue differ. Because the parity eigenvalue
of spin-1 particles is $\sigma $-dependent, there is no zero-mode solution
for spin-1 bosons. Except for the zero-mode solution for spin-0 particles,
the bound-state spectra of spin-1 particles and spin-0 particles are
identical.

\section{Concluding remarks}

The stationary states of spin-1 bosons interacting via a nonminimal vector
coupling was investigated by a technique which maps the DKP equation into a
Sturm-Liouville problem for the physical components of the DKP spinor.
Scattering states in a double-step potential were analyzed and an
oscillatory transmission coefficient was found. An interesting feature of
the scattering is that the transmission coefficient never vanishes, no
matter how large $|V_{0}|$ may be. It was shown that, for a potential of
sufficient intensity, the transmission amplitude exhibits complex poles
corresponding to bound-state solutions. The eigenenergies for bound states
are solutions of transcendental equations classified as eigenvalues of the
parity operator. In the nonrelativistic limit, resonance transmission and
bound-state solutions occur only if the separation of the steps in the
double-step function is very large compared to the Compton wavelength of the
boson ($a\gg 1/m$). The case of a sign potential was analyzed by a limiting
process ($a\rightarrow 0$). In that last case we obtained a non-oscillatory
transmission coefficient and no bound state. As the potential is a double
step, one should not expect the existence of bound states, and it follows
that such bound states are consequence of the peculiar coupling in the DKP
equation.

For a better understanding of those unexpected results, it can be observed
that the first line of (\ref{spin1-ti}) with the double-step potential can
also be written as
\begin{equation}
\left[ -\frac{1}{2m}\frac{d^{2}}{dx^{2}}+V_{\mathtt{eff}}^{\left( \sigma
\right) }\left( x\right) \right] \phi _{I}^{\left( \sigma \right) }\,=E_{%
\mathtt{eff}}\phi _{I}^{\left( \sigma \right) }  \label{Heff}
\end{equation}%
with%
\begin{equation}
V_{\mathtt{eff}}^{\left( \sigma \right) }\left( x\right) =\frac{V_{0}^{2}}{2m%
}\left[ \theta \left( x-a\right) +\theta \left( -x-a\right) \right] -\frac{%
\sigma V_{0}}{2m}\left[ \delta \left( x-a\right) +\delta \left( x+a\right) %
\right]  \label{veff}
\end{equation}%
and%
\begin{equation}
E_{\mathtt{eff}}=\frac{E^{2}-m^{2}}{2m}  \label{sc2}
\end{equation}%
The set (\ref{Heff})-(\ref{sc2}) plus $\int_{-\infty }^{+\infty }dx\,|\phi
_{I}^{\left( \sigma \right) }|^{2}<\infty $ correspond to the
nonrelativistic description of a particle of mass $m$ with energy $E_{%
\mathtt{eff}}$ subject to a potential $V_{\mathtt{eff}}^{\left( \sigma
\right) }$. Therefore one has to search for solutions of the Schr\"{o}dinger
equation for a particle under the influence of a finite square well
potential with attractive (repulsive) $\delta $-functions when $\sigma
V_{0}>0$ ($\sigma V_{0}<0$) situated at the borders. Whether $\sigma V_{0}$
is positive or negative, the effective potential $V_{\mathtt{eff}}^{\left(
\sigma \right) }$ has also a form that would make allowance for bound-state
solutions with $E_{\mathtt{eff}}<V_{0}^{2}/\left( 2m\right) $, and $E_{%
\mathtt{eff}}>0$ if $\sigma V_{0}<0$. For $a\rightarrow 0$, the case of a
sign potential, the effective potential becomes the shifted $\delta $%
-function potential:%
\begin{equation}
V_{\mathtt{eff}}^{\left( \sigma \right) }\left( x\right) \underset{%
a\rightarrow 0}{\longrightarrow }\frac{V_{0}^{2}}{2m}-\frac{\sigma V_{0}}{m}%
\delta \left( x\right)
\end{equation}%
which leads to a non-oscillatory transmission coefficient independent of the
sign of $\sigma V_{0}$, but only for $\sigma V_{0}>0$ there would be one
bound-state solution with a vanishing effective energy ($|E|=m$) independent
of the size of $|V_{0}|$.

Comparison with the results in the Appendix, shows that $\phi _{I}^{\left(
+\right) }$ (in the vector sector) obeys the same equation obeyed by the
first component of the DKP spinor in the scalar sector and, taking account
of (\ref{CUR2}) and (\ref{corrente4}), $\phi _{I}^{\left( +\right) }$ and $%
\phi _{1}$ are square-integrable functions. The space component of the
current ($J^{1}$) for longitudinally polarized spin-1 bosons has the same
form as that one for spin-0 bosons and so the transmission coefficient does
not depend on the spinorial structure, being identical to the transmission
coefficient of spin-0 bosons. There is only one independent component of the
DKP spinor for the spin-0 sector instead of the three required for the
spin-1 sector, and the presence of a space component of the potential may
compromise the existence of bounded solutions for spin-1 bosons when
compared to the solutions for spin-0 bosons with the very same potentials.
As a matter of fact, apart from the zero-mode solution for spin-0 bosons,
the spectrum does not depend on the spinorial structure. The inexistence of
a zero-mode solution for spin-1 boson can visualized from (\ref{veff}). Due
of the presence of the term $\sigma V_{0}$ in $V_{\mathtt{eff}}^{\left(
\sigma \right) }$, one can see that it is possible to trap a spinless boson
with $E_{\mathtt{eff}}=0$ if $V_{0}>0$, but not a spin-1 boson because it is
impossible to satisfy the condition $\sigma V_{0}>0$ for both helicity
quantum numbers.

\vspace{5in}

\noindent {\textbf{Acknowledgments}}

This work was supported in part by means of funds provided by Coordena\c{c}%
\~{a}o de Aperfei\c{c}oamento de Pessoal de N\'{\i}vel Superior (CAPES) and
Conselho Nacional de Desenvolvimento Cient\'{\i}fico e Tecnol\'{o}gico
(CNPq). We thank an anonymous referee for valueble suggestions.

\newpage

\noindent {\large {\textbf{Appendix}}}

For the case of spin 0, we use the representation for the $\beta ^{\mu }$\
matrices given by \cite{ned1}%
\begin{equation}
\beta ^{0}=%
\begin{pmatrix}
\theta & \overline{0} \\
\overline{0}^{T} & \mathbf{0}%
\end{pmatrix}%
,\quad \beta ^{i}=%
\begin{pmatrix}
\widetilde{0} & \rho _{i} \\
-\rho _{i}^{T} & \mathbf{0}%
\end{pmatrix}%
,\quad i=1,2,3  \label{rep}
\end{equation}%
\noindent where%
\begin{eqnarray}
\ \theta &=&%
\begin{pmatrix}
0 & 1 \\
1 & 0%
\end{pmatrix}%
,\quad \rho _{1}=%
\begin{pmatrix}
-1 & 0 & 0 \\
0 & 0 & 0%
\end{pmatrix}
\notag \\
&&  \label{rep2} \\
\rho _{2} &=&%
\begin{pmatrix}
0 & -1 & 0 \\
0 & 0 & 0%
\end{pmatrix}%
,\quad \rho _{3}=%
\begin{pmatrix}
0 & 0 & -1 \\
0 & 0 & 0%
\end{pmatrix}
\notag
\end{eqnarray}%
\noindent $\overline{0}$, $\widetilde{0}$ and $\mathbf{0}$ are 2$\times $3, 2%
$\times $2 and 3$\times $3 zero matrices, respectively, while the
superscript T designates matrix transposition. The spin matrix is given by%
\begin{equation}
S_{i}=%
\begin{pmatrix}
\widetilde{0} & \overline{0} \\
\overline{0}^{T} & s_{i}%
\end{pmatrix}%
,\quad i=1,2,3  \label{s0}
\end{equation}%
where $s_{i}$ are the 3$\times $3 spin-1 matrices $\left( s_{i}\right)
_{jk}=-i\varepsilon _{ijk}$. Because $\beta ^{0}S_{i}=0$, one has $%
\left\langle S_{i}\right\rangle =\left\langle \vec{S}^{\,2}\right\rangle =0$%
, as it should be. Here $P=\left( \beta ^{\mu }\beta _{\mu }-1\right) /3=%
\mathrm{diag}\,(1,0,0,0,0)$ \cite{gue}. The five-component spinor can be
written as $\psi ^{T}=\left( \psi _{1},...,\psi _{5}\right) $ in such a way
that $\psi _{3}$, $\psi _{4}$ and $\psi _{5}$ are the superfluous components
of the DKP spinor ($\psi _{4}=\psi _{5}=0$ in the case of an one-dimensional
movement). In the time-independent case with $p_{y}=p_{z}=0$, the DKP
equation transmutes into%
\begin{equation*}
\left( \frac{d^{2}}{dx^{2}}+k^{2}\right) \phi _{1}=0
\end{equation*}%
\begin{equation}
\phi _{2}=\frac{E}{m}\,\phi _{1}  \label{dkp4}
\end{equation}%
\begin{equation*}
\phi _{3}=\frac{i}{m}\left( \frac{d}{dx}+A_{1}\right) \phi _{1},\quad \phi
_{4}=\phi _{5}=0
\end{equation*}%
where%
\begin{equation}
k^{2}=E^{2}-m^{2}-A_{1}^{2}+\frac{dA_{1}}{dx}  \label{k}
\end{equation}%
Meanwhile,
\begin{equation}
J^{0}=\frac{E}{m}\,|\phi _{1}|^{2},\quad J^{1}=\frac{1}{m}\,\text{Im}\left(
\phi _{1}^{\ast }\,\frac{d\phi _{1}}{dx}\right) ,\quad J^{2}=J^{3}=0
\label{corrente4}
\end{equation}

\end{document}